\documentclass[twocolumn,preprintnumbers,msmath,amssymb,prl,floatfix,final]{revtex4}

\usepackage{graphicx}
\usepackage{dcolumn}% Align table columns on decimal point
\usepackage{bm}% bold math
\usepackage[latin1]{inputenc}
\usepackage[latin1]{inputenc}
\usepackage{verbatim}
\usepackage{float}
\usepackage{color}
\usepackage{amsmath}

\usepackage[]{placeins}

\graphicspath{{./figures/}{./eps/}{./pdf/}}

\begin{document}

%%%%%%%%%%%%%%%%%%%%%%%%%%%%%%%%%%%%%%%%%%%%%%%%%%%%%%%%%%%%%
\title{Click chemistry in ultra-high vacuum - tetrazine coupling with methyl enol ether covalently linked to Si(001)}
%%%%%%%%%%%%%%%%%%%%%%%%%%%%%%%%%%%%%%%%%%%%%%%%%%%%%%%%%%%%%

\author{T. Glaser$^{1}$, J.~Meinecke$^{2}$, L. Freund$^{1}$, C.~L\"anger$^{1}$, J.-N. Luy$^{2,3,\dagger}$, R. Tonner$^{2,3,\dagger}$,  U.~Koert$^{2}$,
and M.~D\"urr$^{1,*}$}
\address{
$^1$Institut f{\"u}r Angewandte Physik and Zentrum f\"ur Materialforschung, Justus-Liebig-Universit\"at Giessen, Heinrich-Buff-Ring 16, D-35392 Giessen, Germany\\
$^{2}$Fachbereich Chemie, Philipps-Universit\"at Marburg, Hans-Meerwein-Stra{\ss}e 4, D-35032 Marburg, Germany\\
$^{3}$Fakult\"at f\"ur Chemie und Pharmazie, Universit\"at Regensburg, Universit\"atsstra{\ss}e 31, D-93053 Regensburg, Germany\\
$^{\dagger}$Current address: Wilhelm-Ostwald-Institut f\"ur Physikalische und Theoretische Chemie, Universit\"at Leipzig, Linn\'estra{\ss}e 2, D-04103 Leipzig, Germany\\
%$^{*}$Email-Address: michael.duerr@ap.physik.uni-giessen.de\\
}

\date{\today}

\bibliographystyle{prsty}
\hyphenation{tempera-ture}

\begin{abstract}
The additive-free tetrazine/enol ether click reaction was performed in ultra-high vacuum (UHV) with an enol ether group covalently linked to a silicon surface: Dimethyl 1,2,4,5-tetrazine-3,6-dicarboxylate molecules were coupled to the enol ether group of a functionalized cyclooctyne which was adsorbed on the silicon (001) surface via the strained triple bond of cyclooctyne. The reaction was observed at a surface temperature of 380~K by means of X-ray photoelectron spectroscopy (XPS). No indications for tetrazine molecules which bind directly to the Si(001) surface via the nitrogen atoms were detected. A moderate energy barrier was deduced for this click reaction in vacuum by means of density functional theory based calculations, in good agreement with the experimental results. This UHV-compatible click reaction thus opens a new, flexible route for synthesizing covalently bound organic architectures.
\end{abstract}

\maketitle

%\newpage

%%%%%%%%%%%%%%%%%%%%%%%%%%%%%%%%%%%%%%%%%%%%%%%%%%%%%
%\section*{INTRODUCTION}
%%%%%%%%%%%%%%%%%%%%%%%%%%%%%%%%%%%%%%%%%%%%%%%%%%%%%

Click reactions \cite{Kolb01AngChem,Sletten09AngChem} are employed in various fields of chemistry such as drug development \cite{Peng17AngChem} and material science \cite{Xi14afm}. When applied in surface chemistry, click reactions may lead to the formation of well-ordered organic layers on inorganic surfaces \cite{Devaraj07qcs,Gouget-Laemmel13jpcc}; moreover, the use of bifunctional molecules in combination with chemoselective reaction schemes could lead to the formation of well-ordered covalently bound organic architectures on inorganic substrates. Semiconductor substrates, in particular the technologically most relevant Si(001) surface, are of high interest for such studies based on their wide range of applications in microelectronics and related technologies.

Single crystal silicon surfaces with high perfection in surface structure and cleanliness are typically prepared in ultra-high vacuum (UHV) \cite{Nedder96rpp}. However, conventional click reactions, such as the widely used copper-catalyzed azide-alkyne cycloaddition \cite{Kolb01AngChem}, require a catalyst dissolved in a solvent and are thus not compatible with a UHV-based approach for multilayer synthesis. Previously reported catalyst-free approaches on surfaces were performed on Cu-containing substrates \cite{Bebensee13jacs,He19ChemMat} and are thus not applicable for a wide range of substrate surfaces.

%%%%%%%%%%%%%%%%%%%%%%%%%%%%%%%%%%%%%%%%%%%%%%%%%%%%%
% Figure: Schema Cyclooctin-enol-ether auf Si(001)
%%%%%%%%%%%%%%%%%%%%%%%%%%%%%%%%%%%%%%%%%%%%%%%%%%%%%
%
\begin{figure}[t!]
	%\vspace{5mm}
	\begin{center}
		\includegraphics[width = 8.5cm]{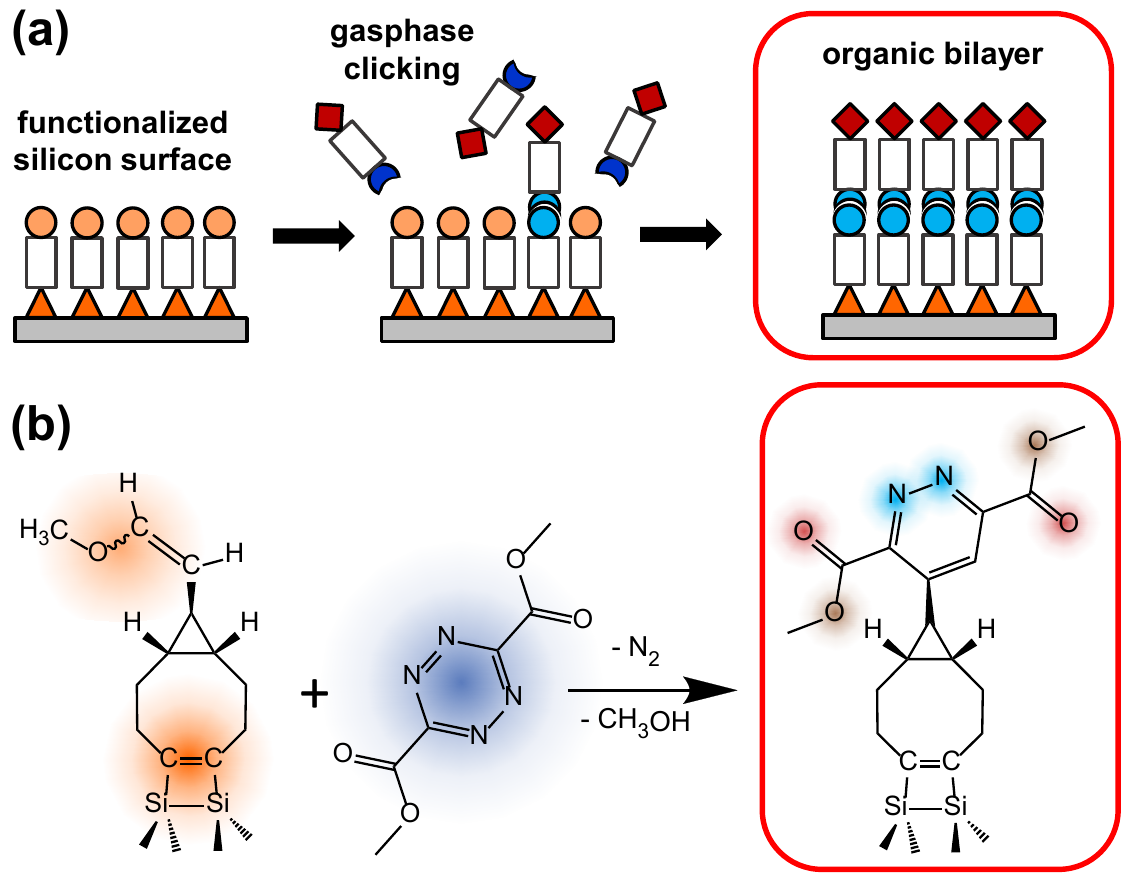}
		\caption[]{(a) Reaction of a bifunctional molecule via gasphase clicking on a first organic layer which was prepared via chemoselective adsorption on the surface of a solid substrate. (b) Methyl enol ether functionalized cyclooctyne (MEECO) adsorbed on Si(001) was further reacted via tetrazine/enol ether coupling which proceeds via abstraction of {N$_2$} and {CH$_3$OH}.
			 \label{fig:Intro}}
	\end{center}
	\vspace{-5mm}
\end{figure}

Here we employ tetrazine/enol-ether coupling, known as additive-free click reaction in solution \cite{Meinecke19OrgLett}, for surface functionalization in ultra-high vacuum. Most recently, substituted cyclooctynes have been shown to form well-ordered monolayers of selectively adsorbed bifunctional molecules on Si(001) \cite{Reutzel16jpcc,Pecher18tca,Laenger19jpcm,Glaser20meeco}, which can be employed for further building up of organic structures (Fig.~\ref{fig:Intro}(a)). With the enol ether group being attached on the surface via a cyclooctyne as a linker (Fig.~\ref{fig:Intro}(b)), we show experimentally that tetrazine/enol-ether coupling can be realized under UHV conditions at moderate temperatures. The results are backed by DFT calculations which find a moderate energy barrier for this reaction in the absence of solvent or catalyst. The tetrazine/enol-ether coupling is independent of the underlying substrate and thus opens a new, flexible route towards covalently bound organic architectures with many applications, e.g., in surface functionalization or organic electronics.

%%%%%%%%%%%%%%%%%%%%%%%%%%%%%%%%%%%%%%%%%%%%%%%%%%%%%
%\section*{METHODS }
%%%%%%%%%%%%%%%%%%%%%%%%%%%%%%%%%%%%%%%%%%%%%%%%%%%%%

The XPS experiments were performed in a UHV chamber with a base pressure $<~1~\times~10^{-10}~$mbar. Si(001) samples were prepared by degassing at 700~K and repeated direct current heating cycles to 1450~K.
A well ordered $2~\times~1$ reconstruction was obtained by cooling rates of about 1~K$/$s \cite{Schwalb07prb, Mette19AngChem}.
The preparation of methyl enol ether functionalized cyclooctyne (MEECO) on Si(001) was carried out according to Ref.~\cite{Glaser20meeco}. MEECO adsorption on Si(001) preferentially takes place via the strained triple bond of the cyclooctyne ring; side reactions include ether cleavage and the formation of an aldehyde group \cite{Glaser20meeco}.

Synthesis of dimethyl 1,2,4,5-tetrazine-3,6-dicarboxylate (short: tetrazine, Fig.~\ref{fig:Intro}(b)) was carried out according to Ref.~\cite{Meinecke19OrgLett}.
Tetrazine was dosed via a leak valve from the vapor phase in a test tube while the sample was kept at constant temperature using direct current heating of the sample and liquid nitrogen cooling of the sample holder.
XPS measurements were performed using an Al K$_{\alpha}$ X-ray source with a monochromator (Omicron XM1000) and a hemispherical energy analyzer (Omicron EA125). All XPS spectra were referenced to the Si~2p$_{3/2}$ peak at 99.4~eV \cite{Reutzel15jpcc}.
Voigt-profiles were used for fitting the data; they are composed of 90~$\%$ Gau$\ss$function and 10~$\%$ Lorentzfunction. If not otherwise stated, full width at half maximum (FWHM) was approximately 0.9~eV in case of the single components of the C~1s and O~1s signals, and approximately 1.5~eV for the N~1s signals; these values are typical for XPS spectra measured in this setup \cite{Laenger18jpcc, Heep20jpcc}.

DFT investigations were done with the Vienna ab initio simulation package (VASP 5.4.4) \cite{Kresse93PRB,Kresse96CMS,Kresse96PRB} and standard PAW-pseudopotentials PBE.54 \cite{co5} with a large core configuration while dispersion effects were considered via the DFT-D3 scheme including an improved damping function \cite{Grimme10jcp,Grimme11cms}. The plane wave energy cutoff was set to 400~eV and a total energy difference of at least $10^{-6}$~eV with ``accurate'' precision was used for SCF convergence. Structural optimizations were performed with the PBE-D3 \cite{Perdew96PRL} exchange correlation functional with the force convergence criterion set to $10^{-2}$~eV/$\mathrm{\AA}$ while more accurate energies were then derived using HSE06-D3 \cite{doi:10.1063/1.2404663} range-separated hybrid functional as single-point energies. For tetrazine, a planar structure has been used, PBE-D3 gives an unphysically buckled structure as minimum. Transition-state structures were calculated with the dimer method \cite{Henkelman99jcp} as implemented in the transition state tools (1.73) for VASP with tighter electronic convergence of $10^{-7}$~eV. For the Si(001) slab calculations, a $\Gamma$-centered $2\times 2\times 1$ $k$-mesh was chosen together with a setup of six layers (two bottom layers frozen and terminated with hydrogen atoms) as determined in previous work \cite{Pecher17chemPhyschem}.  Thermodynamic corrections for the Gibbs energy were calculated at the PBE-D3 level in a pseudo gas phase model by replacing all Si-C bonds with capping hydrogens while keeping the C=C distance fixed. Thus a restricted Hessian calculation is performed except for free molecules (tetrazine, dinitrogen, and methanol) for which the full Hessian is used. Scripts to extract thermodynamic data from the VASP output have been published elsewhere \cite{Heep20jpcc}. In order to improve the energies obtained with HSE06-D3, OSV-PNO-CCSD(T) \cite{C4CP03502J} calculations as implemented in TURBOMOLE~7.3 \cite{TURBOMOLE} were performed for the same gas phase model that was used for the thermodynamic corrections. The higher order correction is then given as follows:
\begin{equation}
\begin{split}
\Delta G^{CCSD(T)}_{slab}=&\Delta E^{HSE06-D3}_{slab}\\
&+\Delta E^{CCSD(T)}_{gas}-\Delta E^{HSE06-D3}_{gas}\\
&+\Delta G^{PBE-D3}_{gas}.
\end{split}
\end{equation}

%%%%%%%%%%%%%%%%%%%%%%%%%%%%%%%%%%%%%%%%%%%%%%%%%%%%%
%\section*{RESULTS and DISCUSSION}
%%%%%%%%%%%%%%%%%%%%%%%%%%%%%%%%%%%%%%%%%%%%%%%%%%%%%

\begin{figure}[t!]
	%\vspace{5mm}
	\begin{center}
		\includegraphics[width = 8.cm]{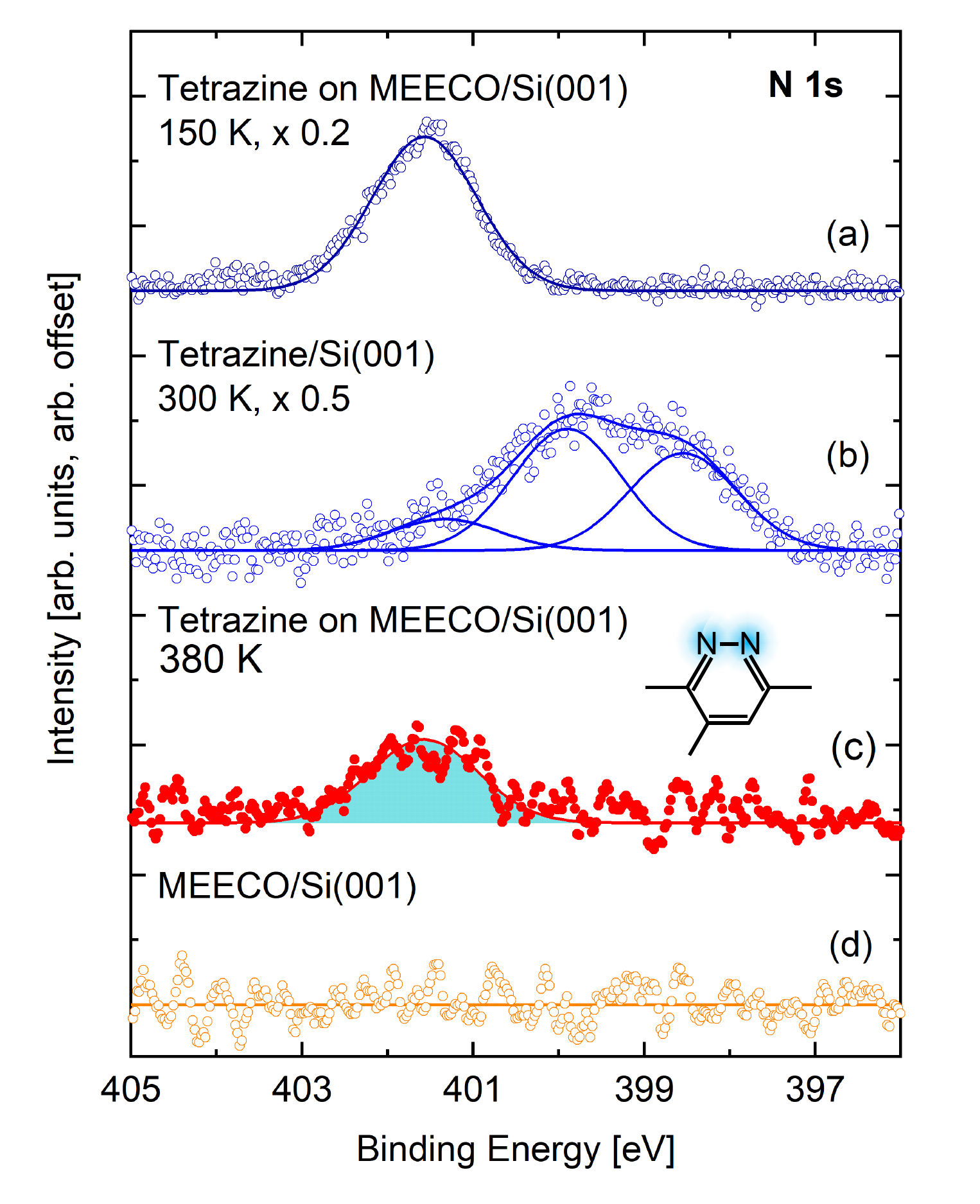}
		\caption[]{N~1s spectra measured at different experimental conditions. (a) Multilayer of tetrazine molecules physically adsorbed on the MEECO covered Si surface at 150~K. In (b), tetrazine was adsorbed on bare Si(001) at 300~K.
In (c), tetrazine was adsorbed on MEECO/Si(001) at 380~K. One single peak is observed which can be assigned to the product of the click reaction of tetrazine on MEECO. For reference, the nitrogen spectrum measured directly after MEECO adsorption is shown in (d), no N signal is identified.
			 \label{fig:N1s}}
	\end{center}
	\vspace{-5mm}
\end{figure}

In Fig.~\ref{fig:N1s}, N~1s spectra of different adsorption experiments are compared. The peak in Fig.~\ref{fig:N1s}(a) results from the nitrogen atoms in a multilayer of tetrazine physically adsorbed on the MEECO-covered Si surface at 150~K. In Fig.~\ref{fig:N1s}(b), adsorption of tetrazine on Si(001) is shown. The peak at 398.5~eV can be assigned to N atoms directly bound to silicon, the peak at 400.0~eV is assigned to further nitrogen atoms in these molecules which are adsorbed on silicon via one or two nitrogen atoms. The small peak at 401.4~eV might be assigned to tetrazine molecules which do not bind via nitrogen atoms but solely via the ester groups to the silicon surface. In contrast to this direct adsorption on silicon, we observe one single peak at 401.6~eV when tetrazine is reacted on the MEECO-covered Si surface at $T_{\rm S} = 380$~K as shown in Fig.~\ref{fig:N1s}(c). We interpret this peak to be the result of the click reaction, as we can exclude a direct binding of tetrazine via the nitrogen atoms to the silicon surface (compare Fig.~\ref{fig:N1s}(b)). As the peak position fits to the position measured for the tetrazine multilayer (Fig.~\ref{fig:N1s}(a)), one might think of some physisorbed tetrazine after adsorption at 380~K as well. However, when tetrazine is adsorbed as multilayer on a MEECO-covered surface at 150~K, no nitrogen signal can be observed when the sample is heated to 300~K (Supporting Information, Fig.~S1). Thus in that case all tetrazine molecules desorb from the surface. This indicates (i) that the signal in Fig.~\ref{fig:N1s}(c) indeed can be assigned to the tetrazine-MEECO coupling and (ii) that this coupling proceeds at temperatures higher than room temperature only.
The similar peak position in Fig.~\ref{fig:N1s}(a) and Fig.~\ref{fig:N1s}(c) can be explained by the fact that the chemical environment of the nitrogen atoms does not change significantly when tetrazine is reacted with MEECO.

%%%%%%%%%%%%%%%%%%%%%%%%%%%%%%%%%%%%%%%%%%%%%%%%%%%%%
% Figure: N 1s overview
%%%%%%%%%%%%%%%%%%%%%%%%%%%%%%%%%%%%%%%%%%%%%%%%%%%%%
%

\begin{figure}[]
	%\vspace{5mm}
	\begin{center}
		\includegraphics[width = 8.cm]{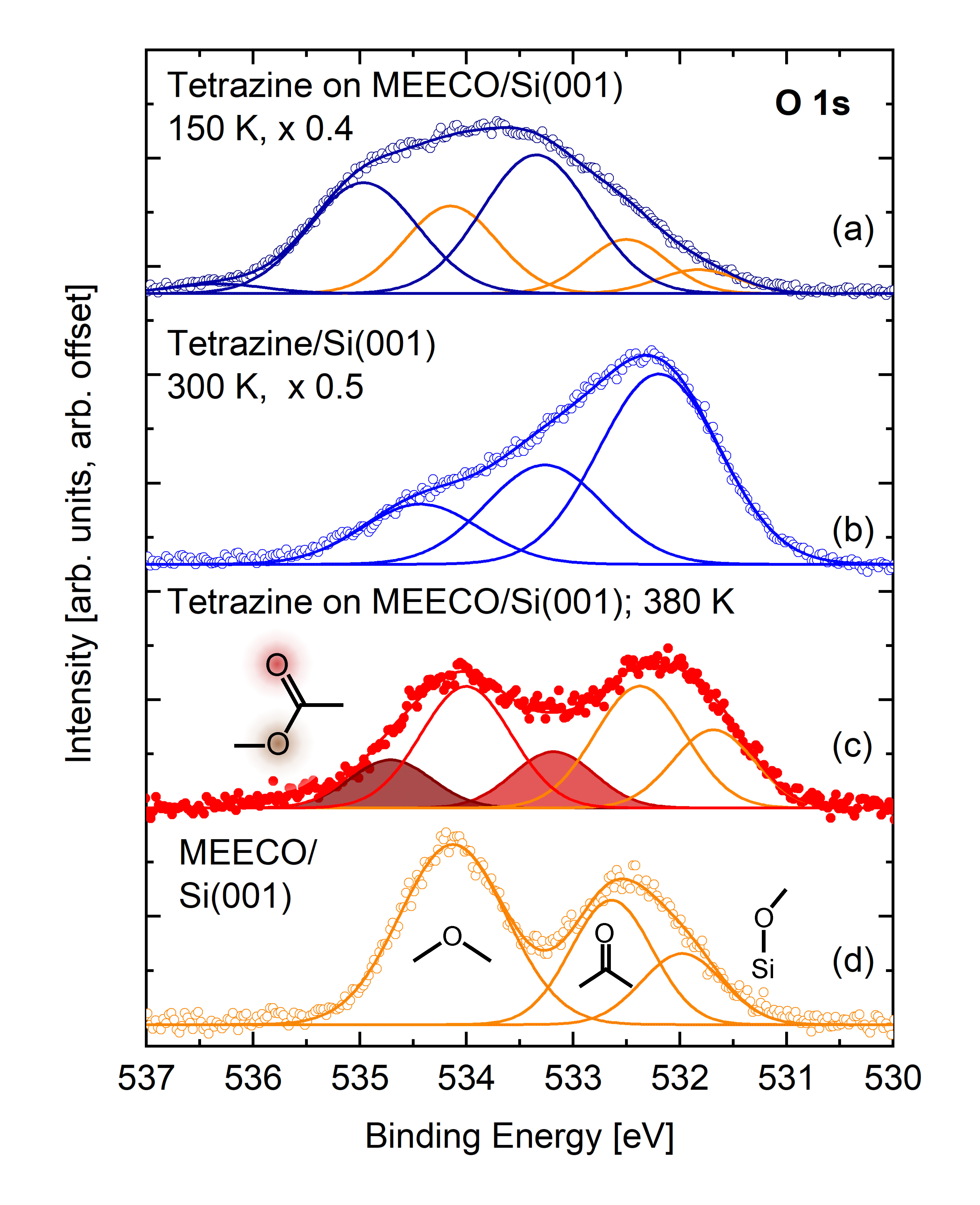}
		\caption[]{O~1s spectra measured at different experimental conditions. In (a), the spectrum of tetrazine on MEECO/Si(001) at 150~K is shown. Most of the intensity (peaks at 533.3~eV and 535.0~eV, blue lines) can be assigned to the two oxygen species present in the tetrazine molecule. Additionally, the oxygen atoms from the adsorbed MEECO molecules contribute to the total intensity (orange components, compare (d), in which the O~1s spectrum after the MEECO adsorption on Si(001) is shown).
In (b), the spectrum of tetrazine adsorbed on bare Si(001) at 300~K is shown. The most intense peak at 532.2~eV can be assigned to O-Si,  indicating that adsorption of tetrazine molecules on Si(001) involves in average more than one oxygen atom per tetrazine molecule. In combination with Fig.~\ref{fig:N1s}(b), multi-tethered molecules including the O and N atoms can be deduced from these experiments.
The spectrum after tetrazine reaction on MEECO/Si(001) at 380~K is shown in (c). The peak of the intact enol ether group decreases; two additional peaks (brown and red) can be assigned to the oxygen atoms in the tetrazine molecule (compare Fig.~\ref{fig:Intro}(b)).

			 \label{fig:O1s}}
	\end{center}
	\vspace{-5mm}
\end{figure}

From the intensity ratio between the N~1s signal after reaction and the O~1s signal of the MEECO-covered surface prior to tetrazine exposure (Fig.~\ref{fig:O1s}(d)), we conclude on approximately 20~\% reacted MEECO molecules on the surface. With a total dose of tetrazine of $10^{-4}$~mbar\,s, this converts into a reaction probability in the order of $10^{-3}$.

%%%%%%%%%%%%%%%%%%%%%%%%%%%%%%%%%%%%%%%%%%%%%%%%%%%%%
% Figure:  O 1s
%%%%%%%%%%%%%%%%%%%%%%%%%%%%%%%%%%%%%%%%%%%%%%%%%%%%%

The O~1s spectra in Fig.~\ref{fig:O1s} were obtained from the same experiments as shown in Fig.~\ref{fig:N1s}.
It has to be taken into account that the adsorbed MEECO on the Si(001) surface already accounts for three peaks in the O~1s spectra (Fig.~\ref{fig:O1s}(d)), as discussed in detail in a previous work \cite{Glaser20meeco}. In brief, these components can be assigned to the intact ether group (534.1~eV) \cite{Glaser20jpcc}, oxygen from the ether group reacted on silicon (532~eV) \cite{Mette14cpc}, and a C=O group as a product of CH$_2$ abstraction (532.7~eV) \cite{ODonnell19jpcc}. In Fig.~\ref{fig:O1s}(c), the spectrum of the coupling product is shown. Two additional peaks are observed, which are assigned to the C-O-C configuration (534.7~eV, brown) and the C=O configuration (533.2~eV, red) in the tetrazine molecule. Both of these configurations have been assigned in the spectrum of MEECO on silicon as well (Fig.~\ref{fig:O1s}(d)), however, with a slightly different binding energy. This difference in binding energy results from two contributions: first, the two components are closely coupled in the ester group of the tetrazine derivative but there is always only one oxygen atom in the configurations related to MEECO on Si(001). Second, the total chemical environment is different: The reacted tetrazine molecule contains two nitrogen atoms, whereas no further heteroatom with an electronegativity higher than for carbon is present in the MEECO molecule.

The peak at 534.0~eV, which is assigned to the intact enol ether group of the MEECO, shows a reduced intensity in Fig.~\ref{fig:O1s}(c) when compared to Fig.~\ref{fig:O1s}(d). This can be seen as a further indication for the click reaction, which reduces the number of intact enol ether groups on the surface.

%%%%%%%%%%%%%%%%%%%%%%%%%%%%%%%%%%%%%%%%%%%%%%%%%%%%%
% Figure: C 1s
%%%%%%%%%%%%%%%%%%%%%%%%%%%%%%%%%%%%%%%%%%%%%%%%%%%%%
%
\begin{figure}[t!]
	%\vspace{5mm}
	\begin{center}
		\includegraphics[width = 8.0cm]{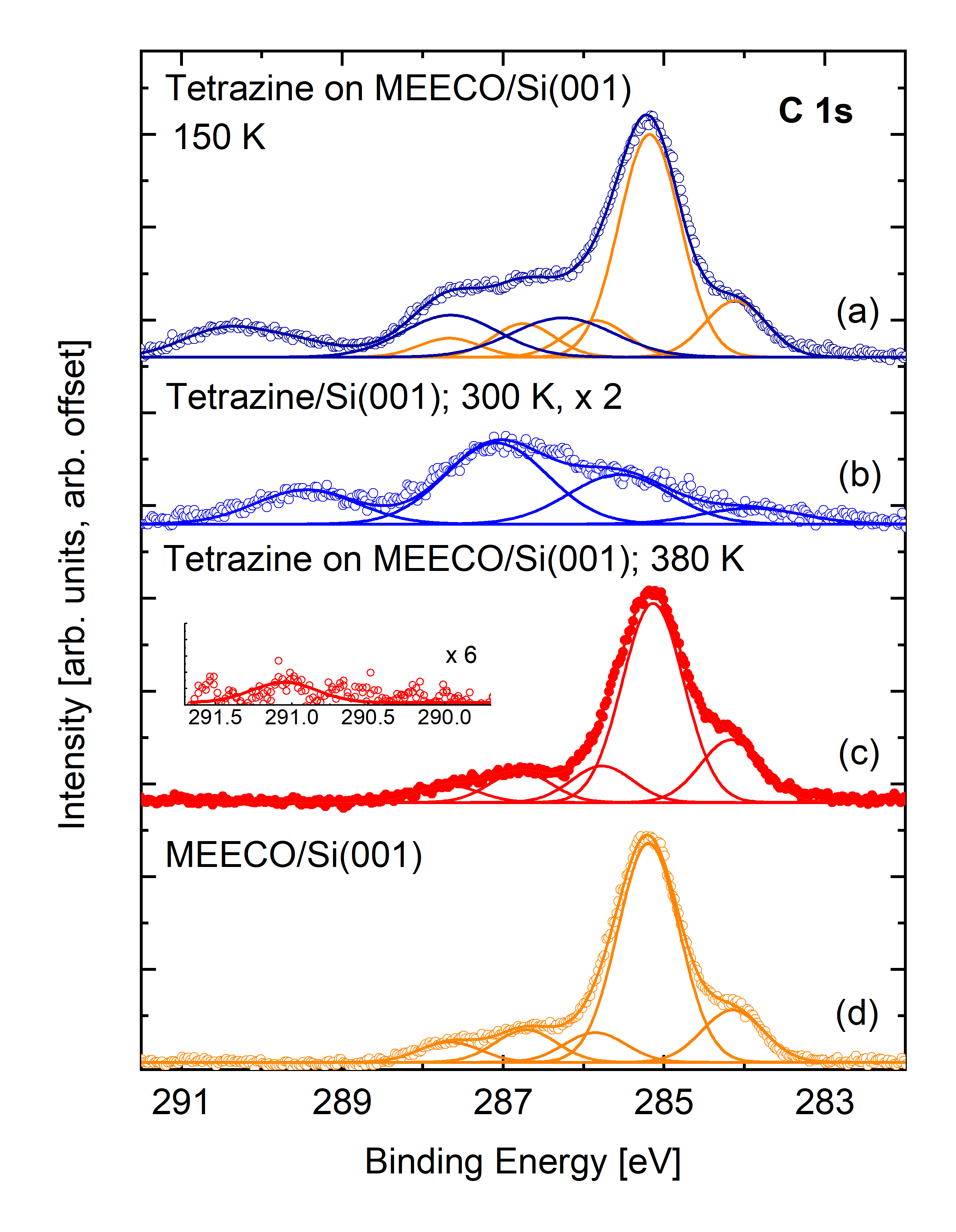}
		\caption[]{C~1s spectra measured at different conditions.
(a) Tetrazine adsorbed on MEECO/Si(001) at 150~K. Blue lines: components attributed to tetrazine, orange lines: components attributed to MEECO on Si(001), compare (d). (b) Tetrazine adsorbed on Si(001) at 300~K. The components attributed in (a) to carbon atoms in tetrazine are shifted to lower binding energy due to adsorption of the molecule to silicon, thus increasing the electronic density also at the carbon atoms being in next neighborhood to the reacting N and O atoms. The spectrum measured after tetrazine reaction on MEECO/Si(001) at 380~K is shown in (c). The inset in (c) indicates a small signal at higher binding energy. In (d), the C 1s spectrum after the MEECO adsorption on Si(001) is shown for comparison.
			 \label{fig:C1s}}
	\end{center}
	\vspace{-5mm}
\end{figure}

The corresponding C~1s spectra are shown in Fig.~\ref{fig:C1s}. All spectra are dominated by the peaks of MEECO adsorbed on silicon (compare Fig.~\ref{fig:C1s}~(d)). In Fig.~\ref{fig:C1s}~(a), when tetrazine is adsorbed on MEECO/Si at 150~K, three further components at relatively high binding energy around 290 to 291~eV and between 286 and 288~eV can be measured. These peaks are assigned to the carbon atoms binding with the oxygen atoms (C-O; O-C=O) and nitrogen atoms in the intact tetrazine molecule. These three components are also present in the spectrum taken after tetrazine adsorption on Si(001) at 300~K ((Fig.~\ref{fig:C1s}~(b)). The spectrum shown in Fig.~\ref{fig:C1s}~(c) was taken after reaction of tetrazine on a MEECO-covered surface at $T_{\rm S}=380$~K. The contribution of the carbon atoms of tetrazine coupled to MEECO to the total carbon signal is low, as the number of reacted MEECO accounts only for approximately one fifth of the MEECO coverage. Nonetheless, we observe a small peak at a binding energy of 291~eV as indicated in the inset of Fig.~\ref{fig:C1s}~(c) which can be assigned to the C atom in the intact ester group (O-C=O) of the tetrazine molecules coupled to MEECO on Si(001).

%%%%%%%%%%%%%%%%%%%%%%%%%%%%%%%%%%%%%%%%%%%%%%%%%%%%%
% Figure: Theory
%%%%%%%%%%%%%%%%%%%%%%%%%%%%%%%%%%%%%%%%%%%%%%%%%%%%%
%
\begin{figure*}[t!]
	%\vspace{5mm}
	\begin{center}
		\includegraphics[width=1.3\columnwidth]{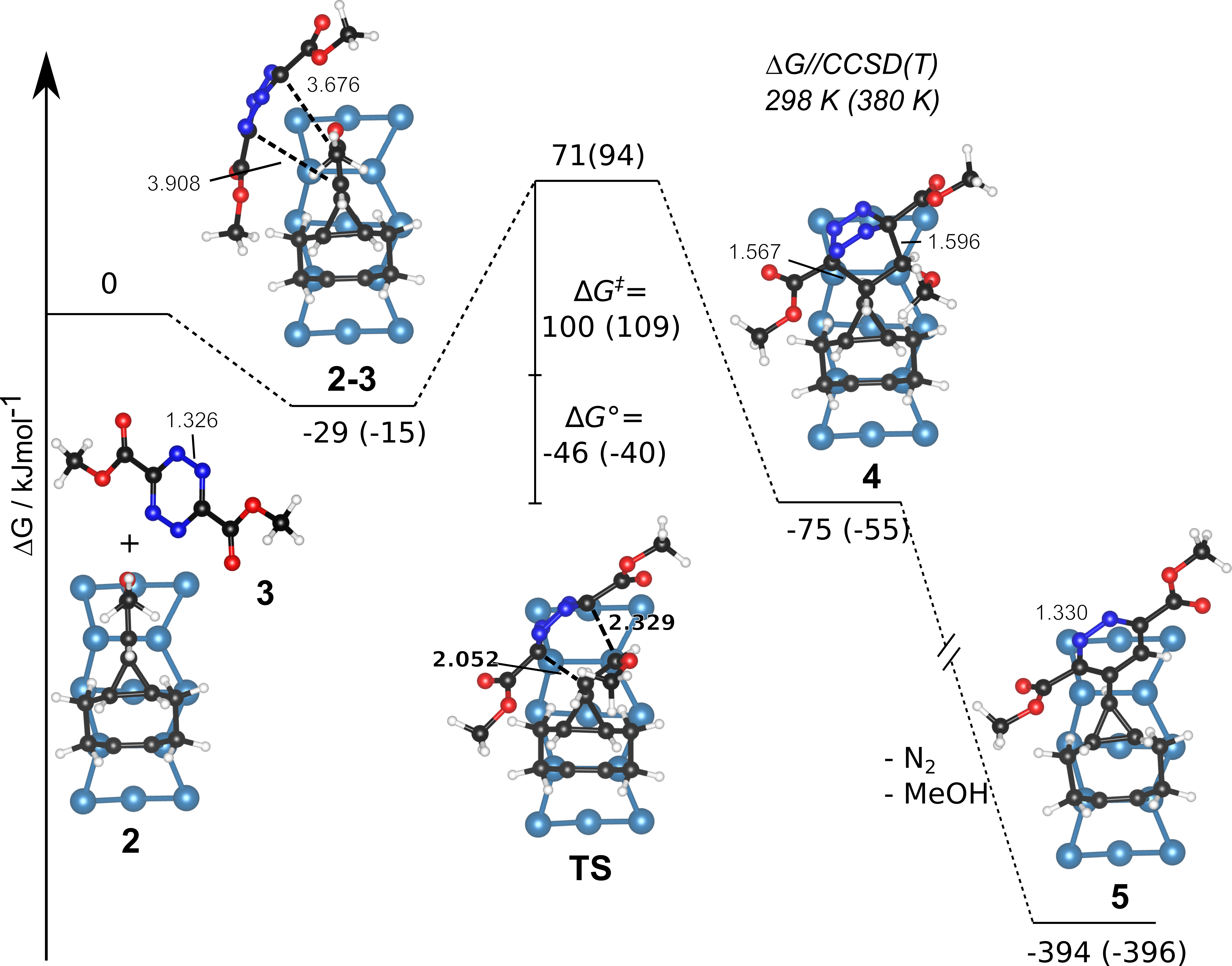}
		\caption[]{Computed reaction pathway (Gibbs free energy $\Delta G$) of tetrazine with MEECO as adsorbed on Si(001). The reaction shows a pre- (\textbf{2-3}) and post-complex (\textbf{4}) connected via a transition state structure (\textbf{TS}) and ends in the product (\textbf{5}) after loosing dinitrogen and methanol. Reaction energies including thermodynamic corrections and high-level energy corrections are given at 298 K and 380 K (in brackets) relative to the separated reactants. Selected bond lengths are given in $\mathrm{\AA}$.
			 \label{fig:theory}}
	\end{center}
	\vspace{-5mm}
\end{figure*}

%%%%%%%%%%%%%%%%%%%%%%%%%%%%%%%%%%%%%%%%%%%%%%%%%%%%%
% Figure: Theory
%%%%%%%%%%%%%%%%%%%%%%%%%%%%%%%%%%%%%%%%%%%%%%%%%%%%%
%
\begin{figure*}[t!]
	%\vspace{5mm}
	\begin{center}
		\includegraphics[width=1.6\columnwidth]{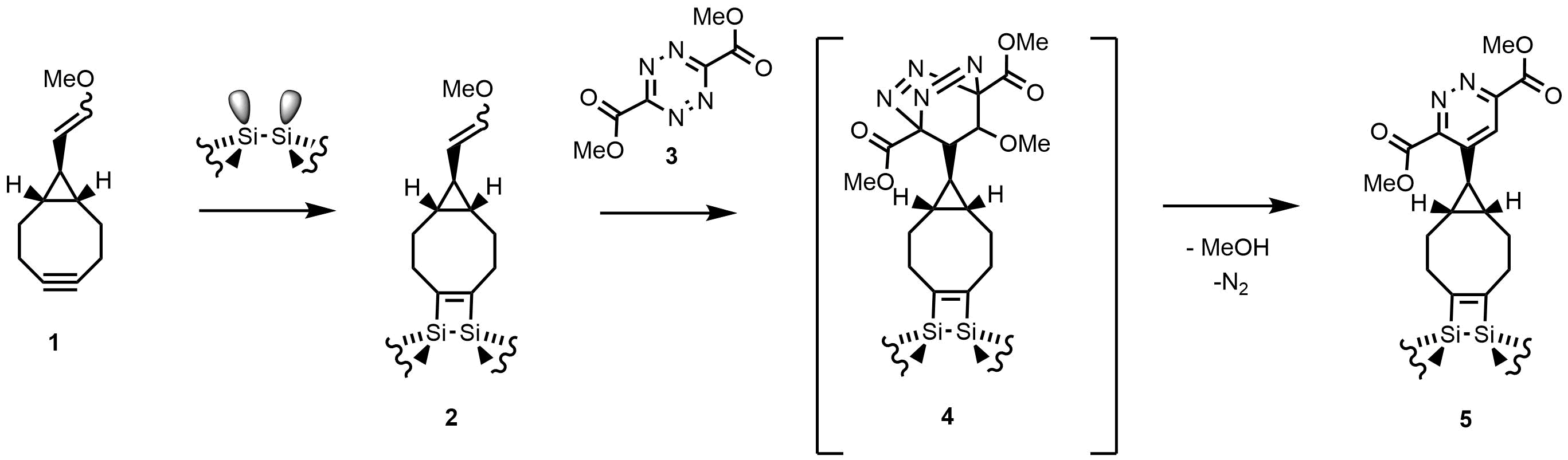}
		\caption[]{Summary of the reaction investigated: MEECO (\textbf{1}) reacts on the dimers of Si(001) via the strained triple bond of cyclooctyne forming \textbf{2} \cite{Glaser20meeco}. 1,2,4,5-tetrazine-3,6-dicarboxylate (\textbf{3}) reacts with \textbf{2} via \textbf{4} to the final product \textbf{5} releasing N$_2$ and MeOH.
			 \label{fig:summary}}
	\end{center}
	\vspace{-5mm}
\end{figure*}

We carried out DFT calculations to shed light on the reaction of adsorbed MEECO with tetrazine with the main results summarized in Fig.~\ref{fig:theory}. We start from intact MEECO adsorbed on Si(001) as investigated before \cite{Glaser20meeco}. The reaction with tetrazine first leads to a physisorbed pre-complex \textbf{2-3} bound by 29 kJ~mol$^{-1}$ at room temperature (all energies refer to Gibbs free energies, $\Delta G$). The reaction then proceeds via a moderate reaction barrier of 100 kJ~mol$^{-1}$ towards the post-complex \textbf{4} which is already 75 kJ~mol$^{-1}$ more stable than the reactants. The final product \textbf{5} is then reached via loosing dinitrogen and methanol. Due to the very high thermodynamic driving force for this second reaction we did not investigate the reaction barrier. The transition state structure \textbf{TS} shows the early state of the formation of two C-C bonds that form the ring structure in the product. The bond lengths are still quite long (2.052 $\mathrm{\AA}$ and 2.329 $\mathrm{\AA}$) in comparison to post-complex \textbf{4} which shows typical values for C-C single bonds. This explains the moderate barrier since the strong deformation in both molecules at the transition state structure is not counterbalanced by stabilization via bond formation processes. The barrier is nevertheless not too high to be overcome at room temperature and perfectly in line with the low reaction rates observed in  experiment. For the higher reaction temperature of 380 K, we find the barrier only mildly increasing (+9 kJ~mol$^{-1}$).

In experiment, the reaction is only observed at elevated temperature while it is not observed when heating the MEECO-covered surface which was prepared with a multilayer of physically bound tetrazine at 150~K (Fig.~S1, Supporting Information). This observation is in agreement with the theoretical results which suggest a substantial conversion rate from \textbf{2-3} to \textbf{4} only at elevated temperatures. When slowly heating the tetrazine-covered surface, desorption via the rather low desorption barrier is favored over reaction via \textbf{TS} with its higher energy barrier.

At this point, we would like to note
%add that there is a strong thermodynamic contribution to the barriers calculated, in particular to the configurations \textbf{A} and \textbf{TS A-B}.
that, in the experiment, the MEECO-covered surface consists of several products and is thus not homogeneous as assumed in the calculations. This may alter the quantitative comparison between experiment and calculations but should have no influence on the qualitative interpretation of the data.

%%%%%%%%%%%%%%%%%%%%%%%%%%%%%%%%%%%%%%%%%%%%%%%%%%%%%
%\section*{CONCLUSION}
%%%%%%%%%%%%%%%%%%%%%%%%%%%%%%%%%%%%%%%%%%%%%%%%%%%%%

As summarized in Fig.~\ref{fig:summary}, coupling between a tetrazine derivative and an enol ether group, the latter being covalently attached on a Si(001) surface via cyclooctyne, has been experimentally observed under ultra-high vacuum conditions, i.e., in the absence of solvent or catalyst. Even under these conditions, the reaction proceeds via a moderate energy barrier between the physisorbed molecule and \textbf{4}, as shown by means of DFT calculations. The further reaction towards the experimentally observed final product \textbf{5} then exhibits a strong thermodynamic driving force. In conclusion, this click chemistry scheme in combination with the chemoselective reactivity of substituted cyclooctynes (\textbf{1} $\rightarrow$ \textbf{2} in Fig.~\ref{fig:summary}) allows for the synthesis of covalently bound molecular architectures even on highly reactive surfaces which require processing in an UHV environment.

%%%%%%%%%%%%%%%%%%%%%%%%%%%%%%%%%%%%%%%%%%%%%%%%%%%%%
\section*{Acknowledgement}
%%%%%%%%%%%%%%%%%%%%%%%%%%%%%%%%%%%%%%%%%%%%%%%%%%%%%

We acknowledge financial support by the Deutsche Forschungsgemeinschaft through SFB~1083 (project-ID 223848855). We thank HRZ Marburg, GOETHE-CSC Frankfurt and HLR Stuttgart for computational resources.

%%%%%%%%%%%%%%%%%%%%%%%%%%%%%%%%%%%%%%%%%%%%%%%%%%%%%
\section*{Associated Content}
%%%%%%%%%%%%%%%%%%%%%%%%%%%%%%%%%%%%%%%%%%%%%%%%%%%%%

Supporting Information include XPS spectra on tetrazine adsorption on MEECO-covered Si(001) after adsorption at 150~K and heating to 300~K as well as information on computational raw data.

%%%%%%%%%%%%%%%%%%%%%%%%%%%%%%%%%%%%%%%%%%%%%%%%%%%%%%%%%%%%%
%\section*{References}
%%%%%%%%%%%%%%%%%%%%%%%%%%%%%%%%%%%%%%%%%%%%%%%%%%%%%%%%%%%%%
\bibliographystyle{prsty}

\bibliography{OrgMolSi,misc2}

\begin{thebibliography}{10}

\bibitem{Kolb01AngChem}
H.~C. Kolb, M.~G. Finn, and K.~B. Sharpless, Angew. Chem. Int. Ed. {\bf 40},
  2004  (2001).

\bibitem{Sletten09AngChem}
E.~M. Sletten and C.~R. Bertozzi, Angew. Chem. Int. Ed. {\bf 48},  6974
  (2009).

\bibitem{Peng17AngChem}
B. Peng, A.-G. Thorsell, T. Karlberg, H. Schgler, and S.~Q. Yao, Angew. Chem.
  Int. Ed. {\bf 56},  248   (2017).

\bibitem{Xi14afm}
W. Xi, T.~F. Scott, C.~J. Kloxin, and C.~N. Bowman, Adv. Funct. Mater. {\bf
  24},  2572   (2014).

\bibitem{Devaraj07qcs}
N.~K. Devaraj and J.~P. Collman, QSAR Comb. Sci. {\bf 26},  1253   (2007).

\bibitem{Gouget-Laemmel13jpcc}
A.~C. Gouget-Laemmel, J. Yang, M.~A. Lodhi, A. Siriwardena, D. Aureau, R.
  Boukherroub, J.-N. Chazalviel, F. Ozanam, and S. Szunerits, J. Phys. Chem. C
  {\bf 117},  368   (2013).

\bibitem{Nedder96rpp}
H. Neddermeyer, Rep. Prog. Phys. {\bf 59},  701  (1996).

\bibitem{Bebensee13jacs}
F. Bebensee, C. Bombis, S.-R. Vadapoo, J.~R. Cramer, F. Besenbacher, K.~V.
  Gothelf, and T.~R. Linderoth, J. Am. Chem. Soc. {\bf 135},  2136   (2013).

\bibitem{He19ChemMat}
C. He, R. Janzen, S. Bai, and A.~V. Teplyakov, Chem. Mater. {\bf 31},  2068
  (2019).

\bibitem{Meinecke19OrgLett}
J. Meinecke and U. Koert, Org. Lett. {\bf 21},  7609   (2019).

\bibitem{Reutzel16jpcc}
M. Reutzel, N. Münster, M.~A. Lipponer, C. Länger, U. Höfer, U. Koert, and M.
  Dürr, J. Phys. Chem. C {\bf 120},  26284  (2016).

\bibitem{Pecher18tca}
L. Pecher and R. Tonner, Theo. Chem. Acc. {\bf 137},  48  (2018).

\bibitem{Laenger19jpcm}
C. L\"anger, J. Heep, P. Nikodemiak, T. Bohamud, P. Kirsten, U. H\"ofer, U.
  Koert, and M. D\"urr, J. Phys.: Condens. Matter {\bf 31},  034001  (2019).

\bibitem{Glaser20meeco}
T. Glaser, J. Meinecke, C. L{\"a}nger, J.-N. Luy, R. Tonner, U. Koert, and M.
  D\"urr, ChemPhysChem {\bf xx},  yy   ((2020)).

\bibitem{Schwalb07prb}
C.~H. Schwalb, M. Lawrenz, M. D\"urr, and U. H\"ofer, Phys. Rev. B {\bf 75},
  085439  (2007).

\bibitem{Mette19AngChem}
G. Mette, A. Adamkiewicz, M. Reutzel, U. Koert, M. Dürr, and U. Höfer, Angew.
  Chem. Int. Ed. {\bf 58},  3417   (2019).

\bibitem{Reutzel15jpcc}
M. Reutzel, G. Mette, P. Stromberger, U. Koert, M. D\"urr, and U. H\"ofer, J.
  Phys. Chem. C {\bf 119},  6018  (2015).

\bibitem{Laenger18jpcc}
C. L\"anger, T. Bohamud, J. Heep, T. Glaser, M. Reutzel, U. H\"ofer, and M.
  D\"urr, J. Phys. Chem. C {\bf 122},  14756   (2018).

\bibitem{Heep20jpcc}
J. Heep, J.-N. Luy, C. L\"anger, J. Meinecke, U. Koert, R. Tonner, and M.
  D\"urr, J. Phys. Chem. C {\bf 124},  9940   (2020).

\bibitem{Kresse93PRB}
G. Kresse and J. Hafner, Phys. Rev. B {\bf 47},  558  (1993).

\bibitem{Kresse96CMS}
G. Kresse and J. Furthmüller, Comput. Mater. Sci. {\bf 6},  15  (1996).

\bibitem{Kresse96PRB}
G. Kresse and J. Furthmüller, Phys. Rev. B {\bf 54},  169  (1996).

\bibitem{co5}
G. Kresse and D. Joubert, Phys. Rev. B {\bf 59},  1758  (1999).

\bibitem{Grimme10jcp}
S. Grimme, J. Antony, S. Ehrlich, and H. Krieg, J. Chem. Phys. {\bf 132},
  154104  (2010).

\bibitem{Grimme11cms}
S. Grimme, S. Ehrlich, and L. Goerigk, Comput. Mater. Sci. {\bf 32},  1456
  (2011).

\bibitem{Perdew96PRL}
J.~P. Perdew, K. Burke, and M. Ernzerhof, Phys. Rev. Lett. {\bf 77},  3865
  (1996).

\bibitem{doi:10.1063/1.2404663}
A.~V. Krukau, O.~A. Vydrov, A.~F. Izmaylov, and G.~E. Scuseria, J. Chem. Phys.
  {\bf 125},  224106  (2006).

\bibitem{Henkelman99jcp}
G. Henkelman and H. Jonsson, J. Chem. Phys. {\bf 111},  7010  (1999).

\bibitem{Pecher17chemPhyschem}
J. Pecher and R. Tonner, ChemPhysChem {\bf 18},  34  (2017).

\bibitem{C4CP03502J}
G. Schmitz, C. H{\"a}ttig, and D.~P. Tew, Phys. Chem. Chem. Phys. {\bf 16},
  22167   (2014).

\bibitem{TURBOMOLE}
{TURBOMOLE V7.2 2017}, a development of {University of Karlsruhe} and
  {Forschungszentrum Karlsruhe GmbH}, 1989-2007, {TURBOMOLE GmbH}, since 2007;
  available from {\texttt http://www.turbomole.com}.

\bibitem{Glaser20jpcc}
T. Glaser, C. L\"anger, J. Heep, J. Meinecke, M.~G. Silly, U. Koert, and M.
  D\"urr, J. Phys. Chem. C {\bf 124},  22619   (2020).

\bibitem{Mette14cpc}
G. Mette, M. Reutzel, R. Bartholomäus, S. Laref, R. Tonner, M. Dürr, U. Koert,
  and U. Höfer, ChemPhysChem {\bf 15},  3725  (2014).

\bibitem{ODonnell19jpcc}
K.~M. O'Donnell, C. Byron, G. Moore, L. Thomsen, O. Warschkow, A. Teplyakov,
  and S.~R. Schofield, J. Phys. Chem. C {\bf 123},  22239  (2019).

\end{thebibliography}
%%%%%%%%%%%%%%%%%%%%%%%%%%%%%%%%%%%%%%%%%%%%%%%%%%%%%%%%%%%%%

\end{document}

\renewcommand{\thepage}{S\arabic{page}}
	\renewcommand{\thefigure}{S\arabic{figure}}
	\renewcommand{\thetable}{S\arabic{table}}

	\setstretch{1.6}
	\noindent Supporting information for:
	\begin{center}
		\begin{large}
			\textbf{Click chemistry in ultra-high vacuum -- tetrazine coupling with methyl enol ether covalently linked to Si(001)}\\
		\end{large}
		\vspace{3mm}
		\begin{small}
		T. Glaser$^{1}$, J.~Meinecke$^{2}$, L. Freund$^{1}$, C.~L\"anger$^{1}$, J.-N. Luy$^{2,3,\dagger}$, R. Tonner$^{2,3,\dagger}$,  U.~Koert$^{2}$,
and M.~D\"urr$^{1,*}$\\
		
$^1$\emph{Institut f\"ur Angewandte Physik and Zentrum f\"ur Materialforschung,
Justus-Liebig-Universit\"at Giessen, Heinrich-Buff-Ring 16, D-35392 Giessen, Germany}\\
$^2$\emph{Fachbereich Chemie, Philipps-Universit\"at Marburg, Hans-Meerwein-Stra{\ss}e 4, 35032 Marburg, Germany}\\
$^{3}$Fakult\"at f\"ur Chemie und Pharmazie, Universit\"at Regensburg, Universit\"atsstra{\ss}e 31, D-93053 Regensburg, Germany\\
$^{\dagger}$Current address: Wilhelm-Ostwald-Institut f\"ur Physikalische und Theoretische Chemie, Universit\"at Leipzig, Linn\'estra{\ss}e 2, D-04103 Leipzig, Germany\\
$^*$\emph{Corresponding author: michael.duerr@ap.physik.uni-giessen.de}
		\end{small}

		\date{\today}

	\end{center}
	\setstretch{1.3}
	\vspace{30mm}
	This Supporting Information includes
	\begin{itemize}
		\item[\textbf{(I)}] N~1s spectra of tetrazine adsorbed on MEECO-covered Si(001) measured at 150~K and after heating to 300~K (Figs.~S1(a) and (b))
        \item[\textbf{(II)}] Link to computational raw data
		\end{itemize}
	
	\pagebreak
	
	\noindent \textbf{I. N~1s spectra of tetrazine adsorbed on MEECO-covered Si(001)}\\
	\\

	\begin{figure}[h!]
		%\vspace{5mm}
		\begin{center}
		\includegraphics[width = 10cm]{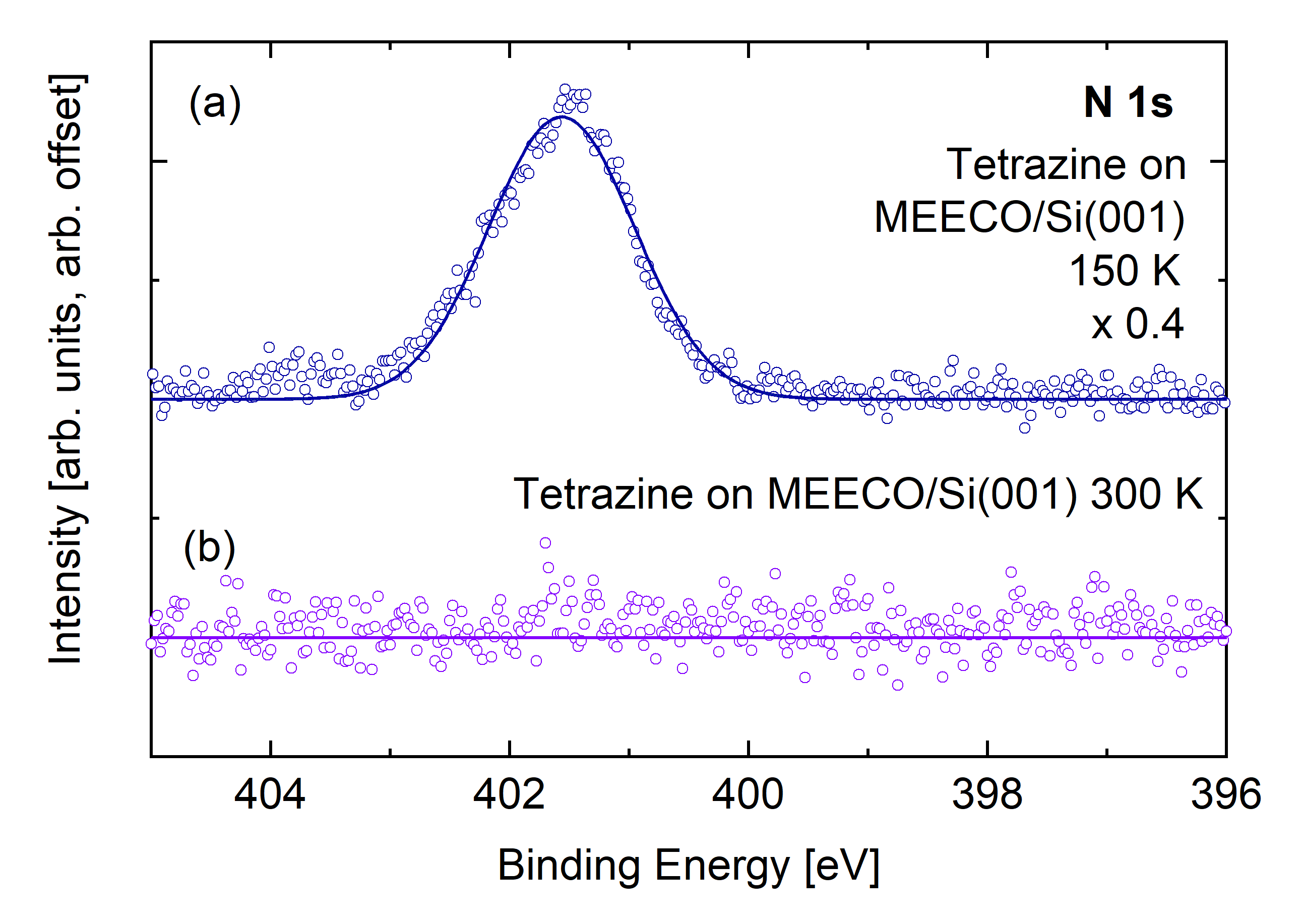}
		\caption[]{(a) N~1s spectra from a tetrazine multilayer adsorbed on a MEECO covered Si(001) surface at 150~K (same experiment as shown in Fig.~2(a) in the main paper). (b) After heating to 300~K (30~min), no N~1s signal is observed any more, all tetrazine molecules are desorbed. No reaction neither with the MEECO nor with the surface has taken place. The latter indicates sufficient passivation by the adsorption of MEECO with respect to direct adsorption of tetrazine on Si(001).\label{fig:Auftau}}
		\end{center}
		%\vspace{-5mm}
	\end{figure}
	
\bigskip

\bigskip

\bigskip

\bigskip

\noindent \textbf{II. Computational raw data}\\

Computational raw data is available at the NOMAD repository:\\ https://dx.doi.org/10.17172/NOMAD/2020.11.19-1